%% file: main.tex
\documentclass[journal]{IEEEtran}
\usepackage{amsmath}
\usepackage{amssymb}
\usepackage{cite}
\usepackage{soul}
\usepackage{color}
\usepackage{graphicx}
\usepackage{url}
\usepackage{multirow}
\usepackage[table]{xcolor}

\usepackage[pscoord]{eso-pic}

\newcommand{\placetextbox}[3]{
  \setbox0=\hbox{#3}
  \AddToShipoutPictureFG*{
    \put(\LenToUnit{#1\paperwidth},\LenToUnit{#2\paperheight}){\vtop{{\null}\makebox[0pt][c]{#3}}}%
  }%
}%

\sethlcolor{white}

\begin{document}

\title{CPES-QSM: A Quantitative Method Towards the Secure Operation of Cyber-Physical Energy Systems}

\author{Juan~Ospina,~\IEEEmembership{Member,~IEEE,} Venkatesh~Venkataramanan,~\IEEEmembership{Member,~IEEE},\\  Charalambos~Konstantinou,~\IEEEmembership{Senior Member,~IEEE}

 \thanks{
 (\textit{Corresponding author}: Juan Ospina)
 
Juan Ospina is with the A-1 Information Systems and
Modeling Group at Los Alamos National Laboratory, Los Alamos, NM, 87545, USA. (e-mail: jjospina@lanl.gov).

Venkatesh Venkataramanan is with the National Renewable Energy Laboratory, Golden, CO, 80401, USA. (email: vvenkata@nrel.gov)

Charalambos Konstantinou is with the Computer, Electrical and Mathematical Sciences and Engineering (CEMSE) Division, King Abdullah University of Science and Technology (KAUST), Thuwal 23955-6900, Saudi Arabia. (e-mail: charalambos.konstantinou@kaust.edu.sa)

}}


\maketitle

\begin{abstract}
Power systems are evolving into cyber-physical energy systems (CPES) mainly due to the integration of modern communication and Internet-of-Things (IoT) devices. CPES security evaluation is challenging since the physical and cyber layers are often not considered holistically. Existing literature focuses on only optimizing the operation of either the physical or cyber layer while ignoring the interactions between them. This paper proposes a metric, the Cyber-Physical Energy System Quantitative Security Metric (CPES-QSM), that quantifies the interaction between the cyber and physical layers across three domains: electrical, cyber-risk, and network topology. A method for incorporating the proposed cyber-metric into operational decisions is also proposed by formulating a cyber-constrained AC optimal power flow (C-ACOPF) that considers the status of all the CPES layers. The C-ACOPF considers the vulnerabilities of physical and cyber networks by incorporating factors such as voltage stability, contingencies, graph-theory, and IoT cyber risks, while using a multi-criteria decision-making technique. Simulation studies are conducted using standard IEEE test systems to evaluate the effectiveness of the proposed metric and the C-ACOPF formulation.
\end{abstract}


\IEEEpeerreviewmaketitle
\input{0-nomenclature}
\input{1-intro}

\input{3-methodcybermetric}

\input{4-methodcyberopf}

\input{5-results}
\input{6-conclusion}

\section*{Acknowledgment}
The authors are thankful to Dr. David Fobes from Los Alamos National Laboratory for his valuable insights and recommendations.

\bibliographystyle{IEEEtran} 
\bibliography{biblio}

\placetextbox{0.6}{0.15}{LA-UR-21-32027}

\end{document}

%% file: 0-nomenclature.tex
\section*{\hl{Nomenclature}}
\label{sect:nomen}
\addcontentsline{toc}{section}{Nomenclature}

\hl{\textit{General Abbreviations}}

\begin{IEEEdescription}[\IEEEusemathlabelsep\IEEEsetlabelwidth{$-----$}]
\item[ACOPF] AC optimal power flow.
\item[ACPF] AC power flow.
\item[C-ACOPF] Cyber-constrained ACOPF.
\item[CPES] Cyber-physical energy systems.
\item[CPES-QSM] Cyber-Physical Energy System Quantitative Security Metric.
\item[CPS] Cyber-physical systems.
\item[CVSS] Common vulnerability scoring system.
\item[DAAs] Data availability attacks.
\item[DIAs] Data integrity attacks.
\item[EPS] Electric power systems.
\item[DERs] Distributed energy resources.
\item[FDPF] Fast decoupled power flow.
\item[HiTL] Human-in-the-loop.
\item[ICT] Information \& communication technologies.
\item[IoT] Internet-of-things.
\item[IT] Information technology.
\item[MCDM] Multi-criteria decision making.
\item[PMU] Phasor measurement units.
\item[RED] Relative electrical distance.
\item[SE] State estimation.
\item[T-ACOPF] Traditional ACOPF.
\end{IEEEdescription}

\hl{\textit{Cyber-Constrained ACOPF}}
\begin{IEEEdescription}[\IEEEusemathlabelsep\IEEEsetlabelwidth{$-----$}]
\item[$N$] Set of buses.
\item[$R$] Set of reference buses.
\item[$G$] Set of generators.
\item[$G_i$] Generator at bus $i$.
\item[$L$] Set of loads.
\item[$L_i$] Load at bus $i$.
\item[$S$] Set of shunts.
\item[$S_i$] Shunt at bus $i$.
\item[$E, E_R$] Set of branches (forward and reverse).
\item[$\Re$] Real part.
\item[$\Im$] Imaginary part.
\item[$v_i^l$] Voltage lower bounds.
\item[$v_i^u$] Voltage upper bounds.
\item[$P^{{g,l}}$] Gen.\ active power lower bounds.
\item[$P^{{g,u}}$] Gen.\ active power upper bounds.
\item[$Q^{{g,l}}$] Gen.\ reactive power lower bounds.
\item[$Q^{{g,u}}$] Gen.\ reactive power upper bounds
\item[$S^{{g,l}}$] Gen.\ complex power lower bounds.
\item[$S^{{g,u}}$] Gen.\ complex power upper bounds.
\item[$c_{2}, c_{1}, c_{0}$] Gen.\ cost components.
\item[$S^{d}$] Load complex power demand.
\item[$Y^s$] Shunt admittance.
\item[$Y, Y^c$] Branch pi-section parameters.
\item[$s_{ij}^{u}$] Complex power flow on line $(i,j)$ upper bounds.
\item[$i_{ij}^{u}$] Current flow on line $(i,j)$ upper bounds.
\item[$\theta_{ij}^{\Delta l}$] Branch angle difference lower bounds.
\item[$\theta_{ij}^{\Delta u}$] Branch angle difference upper bounds.
\item[$\rho$] Threshold value to consider bus `unreliable'.
\item[$\alpha$] Cyber-physical upper bound variable.
\item[$\zeta$] Cyber-physical lower bound variable.
\item[$CQ_k$] CPES-QSM score value at bus $k$.
\item[$\Re({S_{k}^{g}}) = P_{k}^{g} $] Gen.\ $k$ active power output.
\item[$\Im({S_{k}^{g}}) = Q_{k}^{g}$] Gen.\ $k$ reactive power output.
\item[$V_i$] Voltage magnitude at bus $i$.
\item[$\theta_{i}$] Voltage angle at bus $i$.
\item[$S_{ij}$] Complex power flow on line $(i,j)$.
\\
\end{IEEEdescription}

\hl{\textit{CPES-QSM Electrical Factors}}
\begin{IEEEdescription}[\IEEEusemathlabelsep\IEEEsetlabelwidth{$-----$}]
\item[\textit{PI}] Performance Index.
\item[\textit{CRPI}] Contingency Ranking Performance Index.
\item[$P_{l,i}^{^{flow}}$] Power flow on line $l$ with line $i$ out.
\item[$P_l^{^{max}}$] Max. power rating for line $l$.
\item[$n_{PI}$] PI overloaded lines parameter.
\item[\textit{VDI}] Voltage Deviation Index.
\item[\textit{VCPI}] Voltage Collapse Prediction Index.
\item[\textit{SVSI}] Simplified Voltage Stability Index.
\item[$\beta$] SVSI correction factor.
\item[$\Delta V$] Voltage difference.
\end{IEEEdescription}

\hl{\textit{CPES-QSM Graph-Theory Factors}}
\begin{IEEEdescription}[\IEEEusemathlabelsep\IEEEsetlabelwidth{$-----$}]
\item[$\mathcal{G}$] Graph.
\item[$\mathcal{G}_p$] Physical graph.
\item[$\mathcal{G}_c$] Cyber graph.
\item[$\mathcal{N}$] Nodes.
\item[$\mathcal{E}$] Edges.
\item[$s, t, v$] Origin, destination, and evaluated node.
\item[$\mathcal{V}$] Set of all nodes.
\item[$\sigma(s,t)$] Total shortest-paths between $s$ and $t$ .
\item[$d(u,v)$] Shortest-path between node $u$ and $v$.
\item[\textit{BC}] Betweenness Centrality.
\item[\textit{CC}] Closeness Centrality.
\item[\textit{EBC}] Edge Betweenness Centrality.
\end{IEEEdescription}

\hl{\textit{CPES-QSM Cyber Factors}}
\begin{IEEEdescription}[\IEEEusemathlabelsep\IEEEsetlabelwidth{$-----$}]
\item[$P$] Probability.
\item[$I$] Impact.
\item[$AV$] Attack Vector.
\item[$AC$] Attack Complexity.
\item[$UI$] User Interaction.
\item[$PR$] Privileges Required.
\item[\textit{QCR-B}] Quantitative Cyber Risk Base.
\item[\textit{QCR-A}] Quantitative Cyber Risk Attack-Graph.
\end{IEEEdescription}

\hl{\textit{CPES-QSM Computation}}
\begin{IEEEdescription}[\IEEEusemathlabelsep\IEEEsetlabelwidth{$-----$}]
\item[$\nu$] Fuzzy measure.
\item[$\mathfrak{N}$] Set of all criteria.
\item[$\mathfrak{A}, \mathfrak{B}$] Subsets of $N$.
\item[$\emptyset$] Empty set.
\item[$\lambda$] Interaction Index.
\item[\textit{CI}] Choquet Integral.
\end{IEEEdescription}

%% file: 1-intro.tex
\section{Introduction}

\IEEEPARstart{T}{he} modernization and decentralization of electric power systems (EPS) are being facilitated by the integration of distributed energy resources (DERs) and the wide-scale deployment of internet-of-things (IoT) devices and information and communication technologies (ICTs). However, this modernization and transformation from old passive EPS to cyber-physical energy systems (CPES) have their disadvantages. Progressively, CPES are becoming more challenging to secure due to the incorporation of IoT/ICT devices that introduce cyber vulnerabilities to physical systems, thus creating attack vectors not previously considered in traditional power system operations \cite{cpessecurity}.

Recent attack incidents such as BlackEnergy, CrashOverride, and Triton illustrate the growing threat of vulnerabilities in the IoT/ICT infrastructure of power systems \cite{yan, zografopoulos2022distributed}. One prominent example of an attack incident affecting the power grid is the 2015 Ukraine cyber-attack~\cite{jacobs}. In this case, adversaries were able to trip important circuit breakers causing a blackout that affected almost 225,000 customers. Other examples of potential threats to EPS are explored in \cite{keliris2019open, liu2018assessment}, where authors demonstrated how attackers can compromise phasor measurement units (PMUs) by spoofing GPS signals via the use of open-source exploitation methods and open-source intelligence (OSINT) techniques.

Even though the electricity sector has matured in the deployment of protection systems, researchers and stakeholders still struggle when quantifying the cybersecurity status and vulnerabilities of systems operating in a CPES \cite{epricyber}. Conventionally, metrics exist to quantify either the cyber or the physical domain independently. For instance in the cyber domain, Information Technology (IT) systems metrics such as the Common Vulnerability Scoring System (CVSS)~\cite{cvssfirst} exist, which only relies on IT experts' opinions to grade vulnerabilities based on several factors, such as attack vector and attack complexity. While CVSS is useful for analyzing vulnerabilities for IT systems, the framework is insufficient when considering critical infrastructure systems. Trying to address these issues, efforts such as the CERT Resilience Management Model~\cite{cert}, the MITRE's Cyber Resilience Engineering Framework (CREF) \cite{crefmitre}, and the Infrastructure Resilience Analysis Methodology (IRAM) \cite{jacobs}, have been proposed but fail to provide sufficiently useful information to system operators due to not having direct interpretations related to the operation of CPES.

\begin{figure*}
\centering
  \noindent\makebox[\textwidth]{\includegraphics[width=12.0cm]{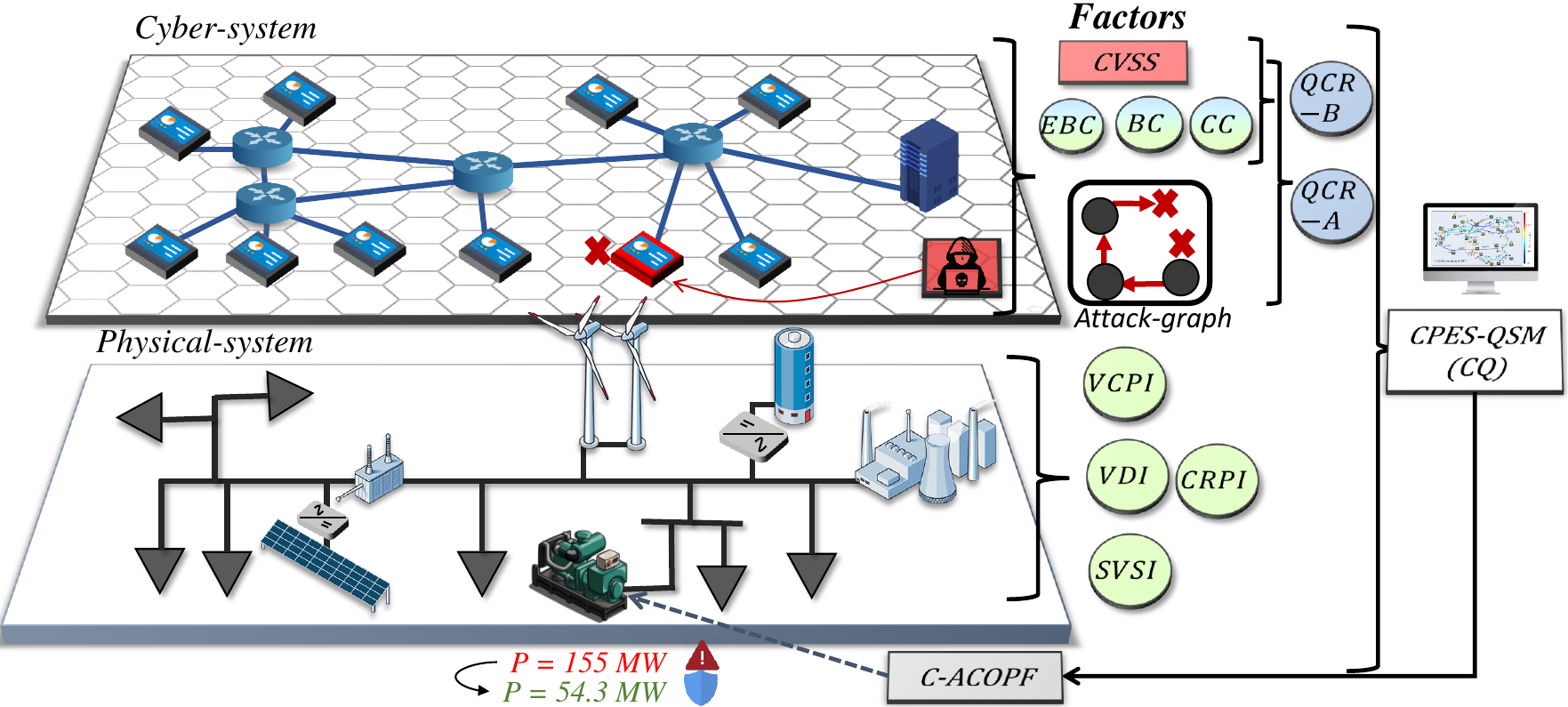}}  \caption{\hl{Overall framework for Cyber-Constrained ACOPF (C-ACOPF) operation based on the Cyber-Physical Energy System Quantitative Security Metric (CPES-QSM).}}
  \vspace{-2mm}
  \label{fig:overallframework}
\end{figure*}

While there are metrics that compute individual physical and cyber resiliency factors, there is a lack of metrics designed to study cyber-physical security rigorously in an integrated manner. Also, many of the metrics proposed in the literature are an aggregation of existing metrics, that need domain specialists to interpret them. That is why authors have focused on addressing this lack of existing metrics by proposing different resiliency metrics and methods specially tailored for small-scale CPES. An example is presented in \cite{choqet1}, where authors propose resiliency metrics that capture the level of preparedness of a distribution system to resist extreme adverse conditions. In \cite{cpindex}, authors propose a stochastic security-oriented risk management technique devised to estimate cyber-physical security indices tailored to measure the security level of a cyber-physical system (CPS). 

The papers presented in \cite{venkataramanan2019cyphyr} and \cite{venkataramanan2019cp} are the ones that are closer to the idea proposed in this paper, since the metrics and methods proposed in these papers follow a similar formulation based on a multi-criteria decision making (MCDM) approach. In these papers, the authors explore the concept of combining different factors, coming from different domains such as the physical, cyber, and device-management domains, using \textit{Choquet Integrals} (CI) to compute an overall resiliency score of the system. The main differences between our proposed approach and the methods presented in these papers are related to the: \textit{(a)} factors considered (different physical, graph-theory, and cyber factors), \textit{(b)} individual scoring metric for each node/bus in the CPES (thus improving visibility and human-in-the-loop (HiTL) processes), and \textit{(c)} the utilization of the proposed cyber-metric as a deciding factor to improve not only the resilience but the overall cyber-physical security state of AC optimal power flow (ACOPF) solutions for CPES.

Another major contribution of this paper is related to the utilization of the proposed cyber-metric as a factor that directly alters the traditional ACOPF solutions of a CPES based on the real-time cybersecurity status of the system, and thus, in a way merging the physical and the cyber domains. A limited number of papers have explored the idea of modifying the ACOPF formulation so that it can integrate the status of the cyber layer of the CPES. One example is the paper presented in \cite{cyberconstrained}, where authors propose a cyber-constrained OPF model for the emergency response of smart grids. The proposed model considers both the physical and cyber network by adding cyber-related constraints to the traditional ACOPF but has the disadvantage of being a `black-and-white' mapping process that assumes the total loss of control of a physical bus mapped to a `cyber-blind' cyber node (cyber-blind meaning a cyber node that becomes invisible due to failing). Another similar paper is the one presented in \cite{valinejad2020community}, where authors develop a multi-agent-based algorithm designed to optimize the power flow of a CPES based on power flow constraints derived from `community resilience' factors such as levels of emotion, empathy, cooperation, and physical health of consumers. The community and cyber-layer factors are measured based on social media sentiment analysis and other social and human factors related to cognitive science and psychology. 

\subsection{Contributions}
Contrasting from the papers examined, this paper aims to address the specific problem of quantitatively measuring  the real-time cyber-physical security of a CPES considering factors that affect both its physical and cyber domains by proposing a novel cyber-metric. It intends to provide an intuitive and easy-to-understand metric that can be used for HiTL operations and can directly reflect the cyber-physical status of the CPES into the ACOPF solutions so that the optimization can be performed not only based on the physical status of the system but considering its current cyber status. The contributions of this paper can be summarized as follows:

\begin{itemize}
    \item A quantitative cyber-physical security metric for CPES is proposed. The Cyber-Physical Energy System Quantitative Security Metric (CPES-QSM) is designed to quantify the current cyber-physical status of every IoT-connected node in a CPES by combining factors from the electrical, IT, and graph-theory domains using an MCDM approach. The metric provides an easy-to-interpret way to evaluate the current state of the CPES.

    \item A cyber-constrained ACOPF formulation that takes into account not only the physical state of the system but also considers the current status of its cyber domain, using the proposed CPES-QSM as a proxy, is proposed. The formulation is intended to restrict the OPF solution space based on the current cyber status of nodes in the system, which can make them `unreliable’ based on vulnerabilities or modeled attack-graphs threats targeting IoT devices deployed in the respective nodes of the CPES.

    \item Experimental case studies are investigated using standard IEEE test systems to demonstrate the usefulness of the proposed cyber-metric and validate the utilization of the cyber-constrained ACOPF formulation for achieving more secure OPF solutions (i.e., considering both the physical and cyber status). The results show how the proposed formulation improves the security and stability of the system when a cyberattack compromises vulnerable nodes.
\end{itemize}

Fig. \ref{fig:overallframework} depicts the overall framework for the use of the proposed CPES-QSM in conjunction with the cyber-constrained ACOPF formulation with the objective of reaching a more secure operating state by adjusting the system's dispatch. The figure illustrates how different factors coming from the cyber, physical, and graph environments are combined into a single quantitative score that is used to improve system's operation.

The rest of the paper is organized as follows. Section II presents the proposed quantitative cyber-physical security metric, CPES-QSM. Section III focuses on presenting the proposed cyber-constrained ACOPF formulation that makes use of the CPES-QSM as a proxy for constraining the optimal solutions based on the cyber factors of the CPES. Section IV presents the experimental setup and case studies performed to evaluate the utility of the proposed cyber-metric and demonstrate its effectiveness for constraining the traditional ACOPF. Finally, Section V presents conclusions and future work.

%% file: 3-methodcybermetric.tex
\vspace{-2mm}
\section{CPES-QSM Computation \& Factors}

This section presents the proposed quantitative cyber-physical metric, CPES-QSM, that is designed to provide a real-time numerical value to the cyber and physical status of an operating CPES. The CPES-QSM cyber-metric provides, via an easy-to-interpret score, operating observability of individual components and/or nodes in a CPES that can be used and interpreted by both human operators and control systems for improving the secure and resilient operation of CPES. 

Before the computation of the CPES-QSM is presented, we need to examine the different factors that affect the final computation of the score. Since our focus is towards CPES, the factors considered cover the following three main domains: (1) \textit{Electrical}, (2) \textit{IT}, and (3) \textit{Graph theory-based}. Within these domains, factors are also classified according to the environment they affect. The environments being considered are: (1) \textit{Physical}, (2) \textit{Cyber}, and (3) \textit{Network}. Each factor, explained in the next subsections, has been categorized using one domain and one environment. Additional details about each factor are presented in Table \ref{tab:allfactors} using a similar format as the one presented by EPRI in \cite{epriformat, epriformat2}. It is important to note that the cyber-metric framework presented in this paper is not tied to the specific factors presented throughout the manuscript and the user/operator has the liberty to choose the factors that he/she thinks better characterizes the physical/cyber/network system being analyzed.

\begin{table*}[t]
\centering
\setlength{\tabcolsep}{1.0pt}
\caption{Details of factors considered for CPES-QSM computation. \\State estimation (SE), Power Flow (PF), Smart-meters (SMs), Phasor Measurement units (PMUs). }
\label{tab:allfactors}
\begin{tabular}{||c|c|c|c|c|c|c|c|c||}
\hline
\textbf{ID} & \textbf{Name} & \textbf{Domain} & \textbf{Environment} & \textbf{Measurement} & \textbf{Formula(s)} & \textbf{Target} & \textbf{Data Source} & \textbf{Report Format} \\ \hline \hline
CRPI & \begin{tabular}[c]{@{}c@{}}Contigency Ranking \\ Performance Index\end{tabular} & Electrical & Physical & Voltages \& Angles & Eq. (\ref{eq:crpi1}) & 0 & SE, PF, IoTs, SMs, PMUs & Decimal \\ \hline
VDI & \begin{tabular}[c]{@{}c@{}}Voltage Deviation \\ Index\end{tabular} & Electrical & Physical & Voltage Magnitude & Eq. (\ref{eq:vdi}) & 0 & SE, PF, IoTs, SMs, PMUs & Decimal \\ \hline
VCPI & \begin{tabular}[c]{@{}c@{}}Voltage Collapse \\ Prediction Index\end{tabular} & Electrical & Physical & Voltages/Admittance matrix & Eq. (\ref{eq:vcpi1}) - (\ref{eq:vcpi2}) & 0 & SE, PF, IoTs, SMs, PMUs & Decimal \\ \hline
SVSI & \begin{tabular}[c]{@{}c@{}}Simplified Voltage \\ Stability Index\end{tabular} & Electrical & Physical & Voltage Phasors & Eq. (\ref{eq:svsi1}) - (\ref{eq:svsi2}) & 0 & SE, PF, IoTs, SMs, PMUs & Decimal \\ \hline
QCR-B & \begin{tabular}[c]{@{}c@{}}Quantitative Cyber Risk \\ Base\end{tabular} & IT & Cyber & CVSS v3.1 Vulnerabilities & Eq. (\ref{eq:qcrbP}) - (\ref{eq:qcrbI}) & $\approx$ 0 & Cybersecurity Assessment & Decimal \\ \hline
QCR-A & \begin{tabular}[c]{@{}c@{}}Quantitative Cyber Risk \\ Attack Graph\end{tabular} & IT & Cyber & CVSS v3.1 Vulnerabilities & Eq. (\ref{eq:qcrbI}) - (17) & $\approx$ 0 & Cybersecurity Assessment & Decimal \\ \hline
BC & Betweenness Centrality & Graph & Network & Topology & Eq. (\ref{eq:bc}) & $\approx$ 0 & Operation Center & Integer \\ \hline
CC & Closeness Centrality & Graph & Network & Topology & Eq. (\ref{eq:cc}) & $\approx$ 0 & Operation Center & Integer \\ \hline
EBC & Edge Betweenness Centrality & Graph & Network & Topology & Eq. (\ref{eq:ebc}) & $\approx$ 0 & Operation Center & Integer \\ \hline
\end{tabular}
  \vspace{-3mm}
\end{table*}

\vspace{-3mm}
\subsection{Electrical Domain Factors}
The \textit{Electrical Domain} factors encompass all factors directly related to the electrical or physical operation of the CPES. 

\subsubsection{Contingency Ranking Performance Index (CRPI)}
\begin{figure}[t]
\centering
\includegraphics[width = 0.38\textwidth]{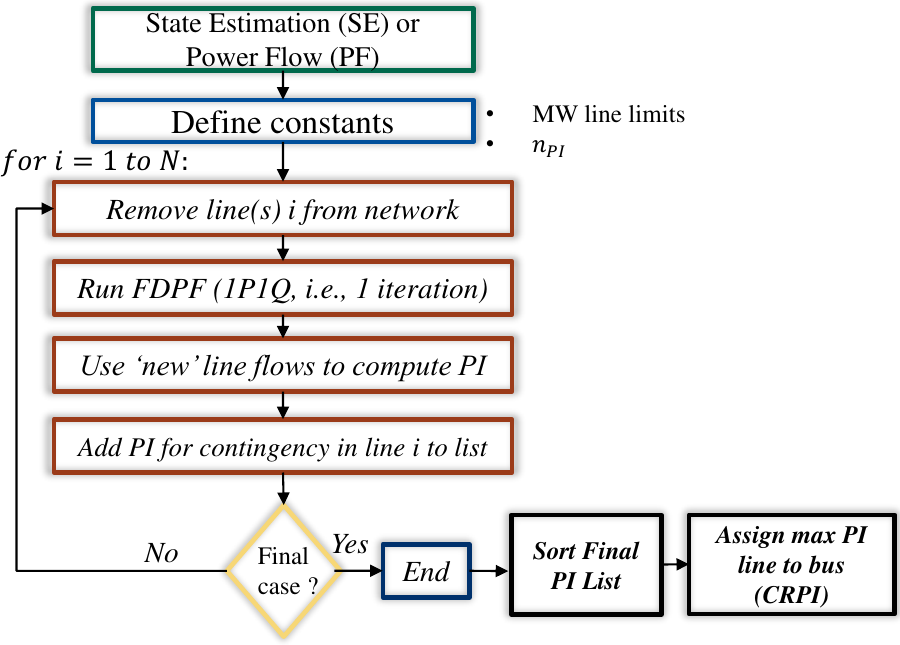}
\caption{\label{fig:crpiprocess} Contingency Ranking Performance Index (CRPI) flowchart process.}
  \vspace{-3mm}
\end{figure}

This factor calculates a contingency performance index, using the fast decoupled power flow (FDPF) \textit{1P1Q} contingency ranking method \cite{wood2013power}, that provides information about the most susceptible lines and buses in a CPES. The specific details for the CRPI factor are shown in Table \ref{tab:allfactors}. The goal of this factor is to calculate the performance index (PI) that tells which lines are most susceptible to overload. Note that \textit{Target} represents the objective value of the factor, i.e., the best value is $0$ while the worst is $1$ (CRPI values are scaled to the [0-1] range). The process for computing the CRPI is based on the process presented 
in \cite{wood2013power}. 
A PI value is assigned to each line outage scenario that can occur in a system. The definition of the PI for a contingency outage $i$ is:
\begin{align}
\label{eq:crpi1}
    PI_{i} = \sum_{l,l \ne i}^{N} \big(\frac{P_{l,i}^{^{flow}}}{P_l^{^{max}}} \big)^{2n_{PI}} \ \ \text{for } i=1,...,N
\end{align}
\noindent where $N$ is the total number of lines in the system, $P_{l,i}^{^{flow}}$ is the power flow on line $l$ with line $i$ out, and $P_l^{^{max}}$ is the maximum power rating for line $l$. The constant $n_{PI}$ is a parameter that allows us to clearly distinguish between overloaded lines and lines with flows within limit. The use of a large value for $n_{PI}$ produces a small $PI$ value (i.e., $\approx 0$) if all line flows are within the limit. On the other hand, if one or more lines are overloaded, the $PI$ value produced will be large. For our test cases, $n_{PI}=2$ is used. The use of the $PI$ value helps us to quickly determine which lines of the system will have a higher impact when taken out, thus which buses are more important in terms of physical contingency security. Using these $PI$ values, a table for each line in the network can be created and sorted from largest to smallest. The values at the top of the list represent the lines that are most important in the contingency ranking; thus they are assigned a higher value. The complete process for computing the CRPI value for each bus in the system is shown in Fig. \ref{fig:crpiprocess}.

As seen in Fig. \ref{fig:crpiprocess}, the process for calculating the $PI$ values for each line in the system starts by obtaining the current status of the system via SE, PF, and/or real-time measurements from in-field IoT devices. Then, using these values alongside the defined constants of the process (i.e., MW limits for lines and $n_{PI}$), the loop in charge of computing the individual $PI$ value for each line in the system is executed. In this loop, line ($i$) is first removed from the network (outage scenario); then an FDPF using the \textit{1P1Q} method is performed in order to estimate the `new' power flows of the system\cite{wood2013power}. The use of the \textit{1P1Q} method, instead of a traditional full ACPF, allows the fast execution of a relatively accurate ACPF without the need of an exact result and multiple iterations, thus reducing the computational complexity and allowing its use in both transmission and distribution systems. After the FDPF is performed, the `new' line flows are computed and used to calculate the corresponding $PI$ value for line $i$ using Eq. (\ref{eq:crpi1}). Finally, the $PI$ value is added to a table or list which will be sorted at the end of the process. After all lines are processed, the final list is sorted from largest to smallest, and the $PI$ value for the line $i_{a,b}$ connecting buses $a$ and $b$ is assigned to buses $a$ and $b$. If a bus is connected to more than one line, the maximum $PI$ of all the lines connected to the bus is selected as the CRPI value for the bus.

\subsubsection{Voltage Deviation Index (VDI)}
This factor calculates the voltage deviation for each bus in the CPES from the nominal 1.0 pu. Similar to CRPI, VDI is considered to be in the \textit{Electrical} domain and the \textit{Physical} environment. The formula to compute the VDI factor is:
\begin{align}
\label{eq:vdi}
    VDI_k = |1.0 - V_{k(in \ pu)}^{mag}|
\end{align}
\noindent where $V_{k(in \ pu)}^{mag}$ is the voltage magnitude measured (in pu) at bus $k$ in the system. The target of the value is 0 since this would mean the bus is exactly at the 1.0 nominal value. This is an easy-to-calculate factor that provides important information related to how far the voltage in a bus is from the nominal 1.0 pu. The specific details for this factor are shown in Table \ref{tab:allfactors}. The goal of this factor is to calculate the voltage deviation (in pu) of a node/bus.

\subsubsection{Voltage Collapse Prediction Index (VCPI)} 
This factor calculates a predicted voltage collapse index, which is designed to determine how close a bus is to voltage collapse. This factor is based on the technique proposed in \cite{balamourougan2004technique}, which is an online technique designed to predict voltage collapse. This factor makes use of voltage magnitudes and voltage angles measured at the respective buses together with the network admittance matrix of the system in order to estimate how close a bus is to voltage collapse when compared to other buses in the system. The method is considered computationally efficient for real-time prediction of voltage collapse in EPS \cite{balamourougan2004technique} since it only needs real-time voltage measurements for its computation. The specific details for the VCPI factor are shown in Table \ref{tab:allfactors}. The goal of this factor is to estimate how close the bus is to voltage collapse and the measurements used are the voltage phasors at sending \& receiving buses, and admittance matrix. The VCPI factor for bus $k$ (receiving bus) is calculated with Eq. (\ref{eq:vcpi1}).
\begin{align}
    \label{eq:vcpi1}
    VCPI_{k} = \bigg|1 - \frac{\sum_{m=1;m \ne k}^{N} V_{m}^{'}}{V_{k}} \bigg|
\end{align}
\noindent where $V_k$ is the voltage measured at bus $k$ and $N$ is the total number of buses in the system. The term $V_{m}^{'}$ is given by:
\begin{align}
    \label{eq:vcpi2}
    V_{m}^{'} = \frac{Y_{km}}{\sum_{j=1;j \ne k}^{N} Y_{kj}} V_m
\end{align}
\noindent where $V_m$ is the voltage phasor at bus $m$ (sending bus), $Y_{km}$ is the admittance between buses $k$ and $m$, and $Y_{kj}$ is the admittance between bus $k$ and $j$. The VCPI factor is computed for each bus in the system.

\subsubsection{Simplified Voltage Stability Index (SVSI)} 
This factor calculates a simplified voltage stability index (SVSI) based on the voltage measurements of the system and load flow results. This factor is based on the indicator presented in \cite{perez2014simplified}, and the relative electrical distance (RED) concept detailed in \cite{yesuratnam2007congestion}. This factor has a similar, but not equal, objective as the VCPI factor presented previously as it is designed to determine how stable a bus in the system is, in terms of voltage collapse. However, the main difference with VCPI is its ability to estimate the voltage stability of a bus based on the RED to the nearest generation bus. In other words, it uses the difference in voltage between the nearest generation bus and the analyzed bus while comparing them to the maximum voltage change in the system (measured considering the substation as the bus with the maximum voltage value). In essence, the estimation of the voltage stability, using this factor, is directly related to how well the generation resources can modify and adjust the voltage at the respective bus. The inputs to compute this factor are: the voltage phasors (voltage magnitude and angle) from the closest generator ($g$), substation ($m$), and receiving ($k$) buses. The admittance matrix is also needed to find the nearest generator to bus $k$, which in turn, can be determined using the RED. The details for the SVSI factor are shown in Table \ref{tab:allfactors}. The SVSI factor for node $k$ can be computed using:
\begin{align}
    \label{eq:svsi1}
    SVSI_{k} = \frac{\Delta V_{k}}{\beta V_k}
\end{align}
\noindent where $V_k$ is the voltage phasor at bus $k$, i.e., our analyzed bus, and $\beta$ is a correction factor that computes the highest difference of voltage magnitudes between any two buses in the system (in the equation, buses $m$ and $l$). The term $\beta$ is computed using:
\begin{align}
    \label{eq:svsi11}
    \beta = 1 - \bigg( max(|V_m| - |V_l|) \bigg)^2
\end{align}
\noindent Finally, the $\Delta V_k$ can be estimated using:
\begin{align}
    \label{eq:svsi2}
    \Delta V_k \cong |V_g - V_k|
\end{align}

\noindent where $V_g$ is the voltage phasor of the nearest generator bus (at bus $g$) to bus $k$. This value is computed using the RED concept that indicates the relative distance of each load bus to all the generator buses in the system \cite{yesuratnam2007congestion, dassios2015unity}.

\vspace{-2mm}
\subsection{Graph-Theory Domain Factors}
\hl{The \textit{Graph-theory Domain} factors encompass all factors that are directly related to the topology of the evaluated network. The topology of the overall CPES is assumed so that the physical and cyber networks are \textit{isomorphic} graphs. This means that each node in the physical-layer of the system is mapped one-to-one with a node in the cyber-layer of the system} \cite{venkataramanan2019cp}. \hl{For instance, if we assume that the physical graph of the system, i.e., the power network, is represented by $\mathcal{G}_p$, while the cyber graph, i.e., the cyber network, is represented by $\mathcal{G}_c$, a bijective relationship can be defined as:}
\begin{align}
    \label{isomorphic}
    f: \mathcal{G}_p \rightarrow \mathcal{G}_c, \ |(x,y) \in \mathcal{G}_p \Rightarrow (a,b) \in \mathcal{G}_c
\end{align}

\begin{figure}[t]
\centering
\includegraphics[width = 0.35\textwidth]{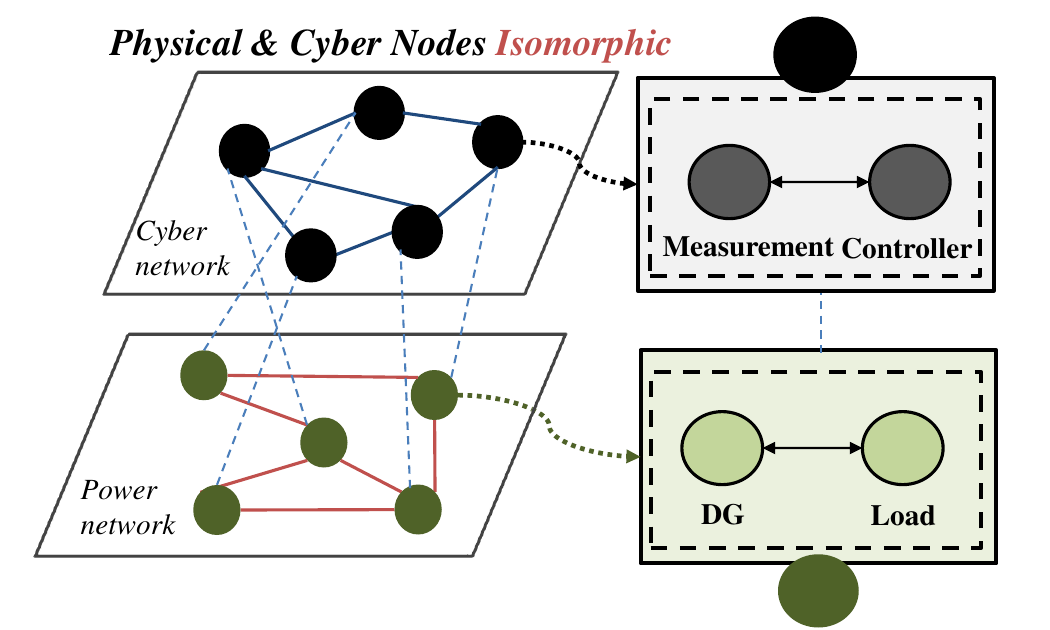}
\vspace{-3mm}
\caption{\label{fig:isomorphfig} Isomorphic physical and cyber network graphs.}
\end{figure}

\hl{It is important to remark that the assumption of `isomorphism' between the cyber and power networks is only considered when mapping the cyber elements with power systems components. This assumption ensures that each physical (power system) node has a representative `mapped' cyber node, which in turn, can be composed of various devices such as multiple measurement and control devices. This mapping allows us to include very detailed cyber models, while retaining the overall structure of the system and thus, lets us determine how a compromised cyber node may directly affect the physical system mapped to it. In other words, isomorphism is only maintained between the primary physical and cyber graphs and minor graphs, composed by multiple devices, can exist inside nodes of these primary graphs.} Fig. \ref{fig:isomorphfig} shows an example of the isomorphic power and cyber graphs. \hl{This assumption is not made for the communication topology between devices, in fact, various types of communication topologies, such as star, ring, or point-to-point, can exist without any discrepancies. }

\hl{Based on the assumptions described above, different graph-theory factors are considered to be included in the computation of the CPES-QSM. These factors are the \emph{betweenness centrality (BC)}, the \emph{closeness centrality (CC)}, and the \emph{edge betweenness centrality (EBC)} of the graph. To calculate these factors, we consider a graph $\mathcal{G}$ composed of $\mathcal{N}$ nodes and $\mathcal{E}$ edges. The order of this graph is given by $\mathcal{G}=(\mathcal{N},\mathcal{E})$.
}

\subsubsection{Betweenness Centrality (BC)}
This factor estimates the importance, i.e., the shortest-path betweenness centrality, of the analyzed node on the overall graph. \hl{It quantifies the number of times a particular node acts as a bridge along the shortest path between two other nodes. The BC of a node $v$ can be calculated as the summation of the fraction of all-pairs shortest paths that pass through $v$:}
\begin{align}
    \label{eq:bc}
    BC(v) = \sum_{s \ne t \ne v\in \mathcal{V}} \frac{\sigma(s,t|v)}{\sigma(s,t)}
\end{align}

\noindent where $s$, $t$, and $v$ are nodes in the set of nodes $\mathcal{V}$. The term $\sigma(s,t)$ is the total number of shortest paths between nodes $s$ and $t$, and the term $\sigma(s,t|v)$ is the number of those paths that pass through node $v$. If $s=t$, then $\sigma(s,t)=1$ and if $v \in s,t$ then $\sigma(s,t|v)=0$ \cite{networkx}.

\subsubsection{Closeness Centrality (CC)}
\hl{This factor estimates the mean distance (geodesic path) from a vertex/node $v$ to every other node. 
The more `central' a node is, the closer it is to all other nodes, thus, the more important in terms of centrality to the overall graph. The CC of node $v$ can be calculated as:}
\begin{align}
    \label{eq:cc}
    CC(v) = \frac{n-1}{\sum_{u=1}^{n-1}d(u,v)}
\end{align}

\noindent where $d(u,v)$ is the shortest-path between node $u$ and $v$, and $n$ is the number of nodes that can be reached from node $v$. A higher value of CC indicates higher centrality \cite{networkx2}.

\subsubsection{Edge Betweenness Centrality (EBC)}
This factor estimates the importance of the edges/lines connected to the analyzed node. The EBC is a factor very similar to BC but calculated for the edges of the graph. The EBC of an edge $e$ can be calculated as the sum of the fraction of all-pairs shortest paths that pass through the edge $e$. This can be calculated as:
\begin{align}
    \label{eq:ebc}
    EBC(e) = \sum_{s \ne t \in \mathcal{V}} \frac{\sigma(s,t|e)}{\sigma(s,t)}
\end{align}

\noindent \hl{where $s$ and $t$ are two arbitrary nodes that exist in the set of nodes $\mathcal{V}$. The term $\sigma(s,t)$ is the total number of shortest paths between nodes $s$ and $t$, and $\sigma(s,t|e)$ is the number of those paths that pass through the edge $e$. The maximum EBC of all the edges connected to a particular node $v$ is the EBC value assigned to that node $v$; since for our purposes, we are mainly interested in characterizing nodes instead of lines.} 
\hl{The details for the BC, CC, and EBC factors are shown in Table} \ref{tab:allfactors}.

\vspace{-2mm}
\subsection{Cyber Domain Factors}

The \textit{Cyber Domain} factors encompass all factors that are directly related to the cyber-layer structure and operation of the CPES. These factors provide a quantitative way of factoring cybersecurity risks that IoT systems introduce into EPS. The processes followed to calculate the quantitative cyber risk base model (QCR-B) factor or the quantitative cyber risk attack graph-based model (QCR-A) factor are inspired from the cyber risk assessment methodology presented in \cite{aksu2017quantitative}. The CVSS v3.1 is used as the scoring system that quantifies the cyber risk of the vulnerabilities that exist in the system \cite{cvssreport}. Essentially, it estimates the difficulty of vulnerability exploitation for each electronic device that exists in a particular node of the cyber-layer of the CPES. The general cyber risk formula used for both QCR-A and QCR-B is $QCR_{B/A} = P \times I$, where $QCR$ is the risk, $P$ is the probability, and $I$ is the impact. The metric values shown in Table \ref{tab:scoresystems} are used to calculate the $P$ variable according to its corresponding calculation process. More details regarding these metrics are given in \cite{deepmetrics}.

\begin{table}[]
\setlength{\tabcolsep}{1.8pt}
\centering
\caption{Exploitability metrics in Common Vulnerability Scoring System Version v3.1 (CVSS v3.1)}
\label{tab:scoresystems}
\begin{tabular}{||c|c|c|c|c||}
\hline
\textbf{\begin{tabular}[c]{@{}c@{}}Score\\ System\end{tabular}} & \textbf{Metric} & \textbf{Abb.} & \textbf{\begin{tabular}[c]{@{}c@{}}Metric\\ Value\end{tabular}} & \textbf{\begin{tabular}[c]{@{}c@{}}Numerical\\ Value\end{tabular}} \\ \hline \hline
\multirow{11}{*}{\textbf{\begin{tabular}[c]{@{}c@{}}CVSS \\ v3.1\end{tabular}}} & \multirow{4}{*}{\begin{tabular}[c]{@{}c@{}}Attack\\ Vector\end{tabular}} & \multirow{4}{*}{AV} & Network & 0.85 \\ \cline{4-5} 
 &  &  & \begin{tabular}[c]{@{}c@{}}Adjacent \\ network\end{tabular} & 0.62 \\ \cline{4-5} 
 &  &  & \begin{tabular}[c]{@{}c@{}}Local \\ network\end{tabular} & 0.55 \\ \cline{4-5} 
 &  &  & Physical & 0.2 \\ \cline{2-5} 
 & \multirow{2}{*}{\begin{tabular}[c]{@{}c@{}}Attack\\ Complexity\end{tabular}} & \multirow{2}{*}{AC} & Low & 0.77 \\ \cline{4-5} 
 &  &  & High & 0.44 \\ \cline{2-5} 
 & \multirow{2}{*}{\begin{tabular}[c]{@{}c@{}}User \\ Interaction\end{tabular}} & \multirow{2}{*}{UI} & None & 0.85 \\ \cline{4-5} 
 &  &  & Required & 0.62 \\ \cline{2-5} 
 & \multirow{3}{*}{\begin{tabular}[c]{@{}c@{}}\\ Privileges \\ Required\end{tabular}} & \multirow{3}{*}{PR} & None & 0.85 \\ \cline{4-5} 
 &  &  & Low & \begin{tabular}[c]{@{}c@{}}0.62 if S = Unchanged\\ 0.68 if S = Changed\end{tabular} \\ \cline{4-5} 
 &  &  & High & \begin{tabular}[c]{@{}c@{}}0.27 if S = Unchanged\\ 0.50 if S = Changed\end{tabular} \\ \hline
 \end{tabular}
   \vspace{-3mm}
\end{table}

\subsubsection{Quantitative Cyber Risk Base Model (QCR-B)}
This factor estimates the cyber risk of a cyber node based on CVSS v3.1 and the vulnerabilities identified in the device that exists in that particular node. This is designed to be the base model of the QCR factor that should be used at nodes that are composed of only one IoT device or access point. The probability $P$ in the risk formulation calculation can be computed as:
\begin{align}
    \label{eq:qcrbP}
    P = AV \times AC \times UI \times PR
\end{align}

\begin{figure}[t]
\centering
\includegraphics[width = 0.42\textwidth]{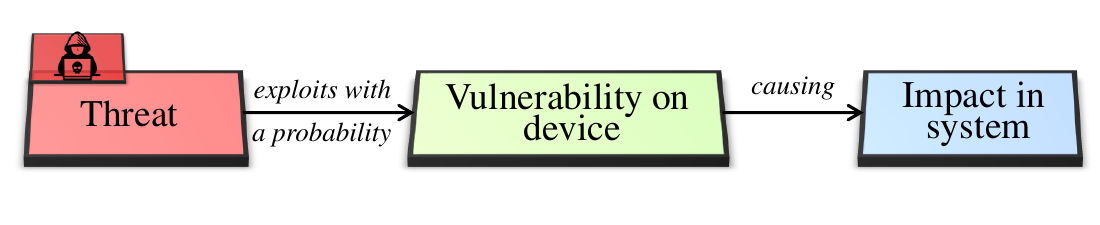}
\vspace{-2mm}
\caption{\label{fig:QCR-B} Quantitative Cyber Risk Base Model (QCR-B) Probability and Impact.}
\vspace{-2mm}
\end{figure}

The impact $I$ of the risk formula is computed using the graph-theory factors presented in the previous subsection times the total bulk impact on the system.
\begin{align}
    \label{eq:qcrbI}
    I = (BC + CC + EBC) \times P^{\%}_{g/l}
\end{align}

\noindent where $P^{\%}_{g/l}$ is the generation or load percentage of the total generation or load in the system. The risk assessment model that describes this calculation process is presented in Fig. \ref{fig:QCR-B}. As seen in this figure, this factor estimates what is the probability or likelihood a vulnerability is exploited on the cyber-physical assets, thus causing an impact on the system. Nodes with high $P$ values (very easy to compromise) and generating/consuming a high percentage of power, with respect to the total load in the system (high $I$ values), are the nodes that are considered more dangerous to the cyber-secure operation of the system.

\subsubsection{Quantitative Cyber Risk Attack Graph-based Model (QCR-A)} 
\begin{figure}[t]
\centering
\includegraphics[width = 0.42\textwidth]{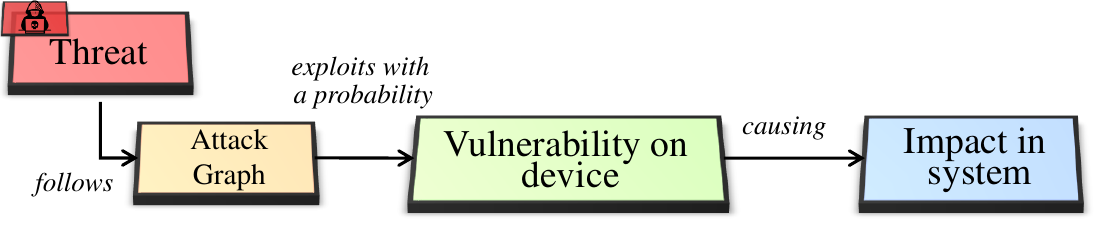}
\vspace{-2mm}
\caption{\label{fig:QCR-A} Quantitative Cyber Risk Attack Graph-based Model (QCR-A) Probability and Impact.}
\vspace{-2mm}
\end{figure}

This factor estimates the cyber risk of a cyber node based on CVSS v3.1 and the vulnerabilities identified in multiple IoT devices that exist in that particular node. This is designed to be the model that should be used at nodes that are composed of multiple electronic devices or access points where \textit{serial} and \textit{parallel} attack graphs can be modeled. Fig. \ref{fig:QCR-A} shows the QCR-A risk assessment process, which differs from the QCR-B process presented previously due to the inclusion of attack graphs that model the path followed by threats that compromise the devices that make up the cyber node.

The probability $P$ in the risk formulation calculation for this version of the factor needs to consider the attack graph method followed to compromise the vulnerabilities in the corresponding devices. Depending on the movement method of the threat, i.e., serial or parallel, the leading probability $P$ is calculated differently. The `leading probability' term is defined as the multiplication of all the probabilities leading to the targeted device. The probability $P$ for a serial movement is calculated by multiplying the leading probabilities with the interim probability at the last device, where the interim probability is defined as the probability of the device calculated from the CVSS scores of the device's vulnerabilities. \hl{The probabilities for a cyber-node with $n$ devices are calculated as follows}:
\begin{align}
\label{eq:probserial}
   P_{n}^{leading} = \prod_{i=1}^{n-1} P_i \\
   P_{n}^{ag} = P_{n}^{leading} \times P_n 
\end{align}
\noindent \hl{where $P_{n}^{ag}$ is the total serial attack-graph probability of a threat that ends up at device \#$n$. $P_n$ is the interim probability of device \#$n$ and $P_{n}^{leading}$ is the leading probability for the path ending at device \#$n$. On the other hand, for a parallel movement of the threat, the total parallel attack-graph probability is calculated by multiplying the parallel probabilities of the leading devices as follows}:
\begin{align}
   P_{n}^{leading} = 1 - \prod_{i=1}^{n-1} (1-P_i) \\
   \label{eq:probparallel}
   P_{n}^{ag} = P_{n}^{leading} \times P_n 
\end{align}

The impact $I$ of QCR-A can be estimated using the same approach used by the QCR-B factor, i.e., Eq. (\ref{eq:qcrbI}). In a real-time application, both of these factors will be mainly affected by the impact $(I)$, which changes according to the percentage of generation or load (at the specific time) with respect to the total load in the system. The specific details for these factors are shown in Table \ref{tab:allfactors}.

\vspace{-3mm}
\subsection{Cyber-Physical Energy System Quantitative Security Metric (CPES-QSM) Computation}

After the required factors are computed, they are combined using an MCDM approach called the \textit{Choquet Integral} (CI) \cite{vu2014choquet}. The main advantage of using the CI for aggregating the factors is that it allows the aggregation of criteria (i.e., factors), $x_n$,  that can have different units, with the objective of providing an overall score. The CI acts as an aggregation function defined with respect to \textit{fuzzy measures}. A fuzzy measure ($\nu$) is defined as a set function that acts on the domain of all possible combinations of a criteria set \cite{meyer2006use}. In other words, fuzzy measures can be thought of as a function that provides a value to every possible combination of inputs. The complexity of the set is determined by the $2^n$ subsets that compose it, where $n$ is the number of criteria, i.e., the number of input factors in the aggregation. 

\hl{Formally, a general fuzzy measure is defined as a set function $\nu:2^n \rightarrow [0,1]$, where $\mathfrak{N} = \{x_1, x_2,...,x_n\}$ represents the set that contains all criteria. The function $\nu$ must be monotonic (i.e., $\nu(\mathfrak{A}) \le \nu(\mathfrak{B})$ whenever $\mathfrak{A} \subset \mathfrak{B}$), and must satisfy $\nu(\emptyset) = 0$ and $\nu(\mathfrak{N}) = 1$, where sets $\mathfrak{A}$ and $\mathfrak{B}$ contain combinations of criteria. Considering that $\mathfrak{A} \subset \mathfrak{B} \subset \mathfrak{N}$, $\nu(\mathfrak{A})$ represents the fuzzy measure or `importance' of the group/subset $\mathfrak{A}$. Fuzzy measures give weights to all possible combinations of criteria, which offers great flexibility when aggregating different types of criteria using the CI for decision-making processes.}

The initial weights (i.e., fuzzy measures) for the individual criteria (e.g., $\nu(\{x_1\}), \nu(\{x_2\})$, etc.) must be assigned by `experts', which are in charge of giving importance to the respective individual factors. This `weighting' process can be performed iteratively using a HiTL approach. On the other hand, the fuzzy measures of the sets that combine multiple factors (e.g., $\nu(\{x_1, x_2\}), \nu(\{x_2, x_3\})$, etc.) are computed using the \textit{interaction index} $\lambda$ estimated as:
\begin{align}
    \label{eq:computelambda}
    \lambda+1 = \prod_{i=1}^{n} (1+\lambda \nu_i), \ \ \text{where } -1 \le \lambda < 0
\end{align}
\noindent The interaction index measures the interaction among
criteria, i.e., it can be considered as a measurement of the interaction between criteria (factors) in the decision-making process. If $\lambda<0$, a correlating relationship between the criteria exists. If $\lambda\approx0$, no relationship between the factors exists, i.e., they contribute independently to the score \cite{vu2014choquet}. 

Using the calculated interaction index ($\lambda$), the fuzzy measures of sets that combine multiple factors are calculated as:
\begin{align}
    \label{eq:computefuzzyneasures}
    \nu(\{x_1,x_2,...,x_n\})=\frac{1}{\lambda} \bigg| \prod_{i=1}^{n} (1+\lambda \nu_i) - 1  \bigg|
\end{align}

Finally, using the calculated fuzzy measures and the values of the factors $x_1,x_2,..., x_n$, the CI can be computed as:
\begin{align}
    \label{eq:choquetintegral}
    CI_{\nu} = \sum_{i=1}^{n} [x_i - x_{i-1}] \nu(F_i)
\end{align}

\noindent where $x_0 = 0$ by convention, and $F_i = {i,...,n}$ is the subset of indexes of the $n-i+1$ largest component. The output of the CI usually lies in the unit interval $[0,1]$; however, other choices can be defined \cite{vu2014choquet}. A simple numerical example of how to compute $\lambda$ and all the fuzzy measures of a set of three criteria (i.e., three factors) is presented below.

Let us assume that there exist a set of three criteria $x_1, x_2, x_3$ that we want to combine using the CI. To compute the CI, we must first compute the fuzzy measures using Eq. (\ref{eq:computefuzzyneasures}) and their corresponding $\lambda$ using Eq. (\ref{eq:computelambda}). To do this, fuzzy measures for the individual inputs $x$ must be assigned. These are assigned as `expert' weights (i.e., fuzzy measure values) given by experts. So, $\nu_0 = 0$,  $\nu_1 = 0.42$,  $\nu_2=0.5$, and $\nu_3=0.62$, where $\nu_0 \equiv \nu(\{\emptyset\})$, $\nu_1 \equiv \nu(\{x_1\})$, $\nu_2 \equiv \nu(\{x_2\})$, and $\nu_3 \equiv \nu(\{x_3\})$. Then, using Eq. (\ref{eq:computelambda}), $\lambda$ can be solved for as $\lambda+1=(1+\lambda \nu_1)(1+\lambda \nu_2)(1+\lambda \nu_3)$. For this particular example, $\lambda = -0.748$. Finally, using the calculated $\lambda$, we can use Eq. (\ref{eq:computefuzzyneasures}) to compute the remaining fuzzy measure values for all the combinations of the set of criteria. The corresponding values calculated for the example given are $\nu(\{x_1, x_2\}) = 0.75$, $\nu(\{x_1, x_3\}) = 0.82$, $\nu(\{x_2, x_3\}) = 0.86$, and $\nu(\{x_1, x_2, x_3\}) = 1.0$. With the fuzzy measure (i.e, weights) calculated, the CI can be then computed for any input values $x_1, x_2$, and $x_3$ using Eq. (\ref{eq:choquetintegral}).

For our case, the CI is the method used to combine the different \textit{Electrical}, \textit{IT}, and \textit{Graph-theory} factors that characterize the current state of a CPES, and thus compute the CPES-QSM for each node that exists in the CPES. For instance, the factors used as inputs to the CI that produces the CPES-QSM for node $i$ could be $x_1^i = CRPI_{i}$, $x_2^i = QCR\text{-}B_{i}$, $x_3^i = VDI_{i}$, $x_4^i = SVSI_{i}$, and $x_5^i = VCPI_{i}$. Each one of these factors is given a corresponding `importance' or `expert' weight (fuzzy measure) $\nu_1,\nu_2,...,\nu_5$. It is important to note that other papers in the literature, such as \cite{venkataramanan2019cp} and \cite{venkataramanan2019cyphyr}, have also used the CI approach for combining different factors in the process of computing cyber-physical resilient analysis metrics, due to the CI characteristics of allowing the aggregation of non-additive nonlinear criteria without assuming independence of each criterion. More details about the main differences between our proposed approach and works in the literature are specified in Section I.

%% file: 4-methodcyberopf.tex
\vspace{-2mm}
\section{Cyber-Constrained ACOPF Formulation}

This section presents an overview of the proposed cyber-constrained ACOPF (C-ACOPF) formulation. Subsection \ref{sect3:a} focuses on presenting and discussing the traditional ACOPF (T-ACOPF) formulation. Subsection \ref{sect3:b} presents the modifications made to the traditional mathematical formulation of the ACOPF so that it can take into account both physical and cyber operating factors of the CPES being optimized; allowing for the consideration of cyber factors in its decision-making process.

\vspace{-3mm}
\subsection{Traditional ACOPF Full Formulation}
\label{sect3:a}

The classical ACOPF formulation can be written as presented in Eqs. (\ref{eq:acopfcostfunc}) - (\ref{eq:acopflast}). Eq. (\ref{eq:acopfcostfunc}) is the objective cost function of the system, Eq. (\ref{eq:slimits}) - (\ref{eq:Sinnodes}) represent the operational power constraints, Eq. (\ref{eq:Sij}) - (\ref{eq:Currentbranchlimit}) represent the branch power and current constraints, and Eq. (\ref{eq:acopflast}) represent the voltage angle difference constraints. There exist other formulations and extensions such as security-constrained optimal power flow, DC OPF, and other relaxations \cite{wang2007computational, wang2007computation, yongfu1, venzke1}. However, our proposed C-ACOPF is based on the traditional nonconvex ACOPF polar formulation presented.

\vspace{-3mm}
\begin{equation}
\label{eq:acopfcostfunc}
   \text{min~~} \sum_{k\in G} c_{2k}(\Re(S_{k}^g))^2 + c_{1k}(\Re(S_{k}^g)) + c_{0k}
\end{equation}

\vspace{-2mm}
\text{S.t.:}
\vspace{-9mm}
\begin{gather}
    \label{eq:silimits}
    \theta_r = 0, \ \forall r \in R \\
    \label{eq:slimits}
    S_k^{gl} \leq S_k^{g} \leq S_k^{gu}, \ \forall k \in G \\
    \label{eq:vlimits}
    v_i^l \leq |V_i| \leq v_i^u, \ \forall i \in N \\
    \label{eq:Sinnodes}
       \begin{aligned}
        \!\sum_{k \in G_i}\!S_k^g\!-\!\sum_{k \in L_i}\!S_{k}^d\!-\!\sum_{k \in S_i}\!(Y_k^s)^*|V_i|^2\!&=\!\!\!\!\!\!\!\!\!\!\sum_{(i,j)\in E_i \cup E_i^R}\!\!\!\!\!\!\!\!\!\!S_{ij}\\
        & , \forall i \in N  
    \end{aligned}\\
    \label{eq:Sij}
    S_{ij} = (Y_{ij} + Y_{ij}^c)^* |V_i|^2 - Y_{ij}^* V_i V_j^*, \ \forall (i,j) \in E \\
    \label{eq:Sji}
    S_{ji} = (Y_{ij} + Y_{ji}^c)^* |V_j|^2 - Y_{ij}^* V_i^* V_j, \ \forall (i,j) \in E \\
    \label{eq:Spowerbranchlimit}
    |S_{ij}| \leq s_{ij}^u, \ \forall (i,j) \in E \cup E_R \\
    \label{eq:Currentbranchlimit}
    |I_{ij}| \leq i_{ij}^u, \ \forall (i,j) \in E \cup E_R \\
    \label{eq:acopflast}
    \theta_{ij}^{\Delta l} \leq (V_i V_j^*) \leq \theta_{ij}^{\Delta u}, \ \forall (i,j) \in E
\end{gather}

\vspace{-4mm}
\subsection{Cyber-Constrained ACOPF Formulation}
\label{sect3:b}

From the formulation presented in the last subsection, it can be observed how the T-ACOPF formulation essentially prioritizes minimizing a cost function that only takes into account the physical states of the system. However, the C-ACOPF formulation presented in this subsection is intended to include the cyber-physical perspective of the system. This perspective is facilitated by using the CPES-QSM, presented in the previous section, as a proxy for the cyber-physical factors operating conditions that make up the CPES. The general formulation in Eq.~\eqref{eq:acopfcostfunc}-\eqref{eq:acopflast} can be stated in a simplified form as:  

\vspace{-5mm}

\begin{gather}
   \text{min~~} C(x) \\
    \text{s.t.~~} G(x) =0 \\
    \label{eq:Hlessl} H(x) \leq l \\
    x_{min} \leq x \leq x_{max}  
\end{gather}

\vspace{-1mm}

\noindent where $C(x)$ is the cost function, $G(x)$ are the equality constraints, and $H(x)$ are the inequality constraints of the problem. The state variables consist of $x$ for a particular generation node $k$. Note that the metric can also be extended to controllable load buses. The CPES-QSM is included in the formulation to bias the optimization solution towards `reliable' generation sources, i.e, the OPF formulation will aim to decrease the power injections from buses where the CPES-QSM value are above a particular threshold. The variable $\rho$ provides the threshold value for which a bus with a CPES-QSM score above this threshold is considered `unreliable'. In practice, this value must be determined by expert operators in charge of managing the system being analyzed and may be different for different systems. Considering the particular threshold defined, the CPES-QSM score ($CQ_k$) directly impacts the upper bounds of the state variables $P_k^g$ and $Q_k^g$ as a scalar multiple. It is important to note that the $CQ_k$ value/score is computed using Eq. (\ref{eq:choquetintegral}), i.e., $CQ_k = CI$. The variables $\zeta$ and $\alpha$ are defined for the lower and upper bounds of the state variables. The new cyber-physical variables $\rho$, $\alpha$, $\zeta$, and $CQ_k$ are included in the inequality constraints of the simplified ACOPF formulation by expanding Eq.~\eqref{eq:Hlessl} as follows,

\vspace{-4mm}
\begin{gather}
\label{eq:35}
      (1-\zeta) P_k^{gl} \leq P_k^{g} \leq \alpha P_k^{gu}, \ \forall k \in G \\  
      (1-\zeta) Q_k^{gl} \leq Q_k^{g} \leq \alpha Q_k^{gu}, \ \forall k \in G \\  
      \label{eq:38}
      \alpha (\rho,\!CQ_k,\!\zeta)\!=\!\begin{cases}
    \!0\!,&\!\text{if }\!CQ_k\!\geq\!\rho\!\text{ for }\!\zeta=1\\
    (\frac{P_k^{gl}}{P_k^{gu}},1),&\!\text{if }\!CQ_k\!\geq\!\rho\!\text{ for }\!\zeta=0\\
    1\!,&\!\text{if }\!CQ_k\!<\!\rho\!\text{ for }\!\zeta=0\\
    \end{cases}
    \end{gather}

The new lower bounds of the state variables become $(1-\zeta) \times P_k^{gl}$ and  $(1-\zeta) \times Q_k^{gl}$. Similarly, the new upper bounds are $\alpha \times P_k^{gu}$ and $\alpha \times Q_k^{gu}$.  The value of $CQ_k$ changes according to the current cyber-physical state of the system and differs according to different types of attacks, e.g., data integrity attacks (DIAs), data availability attacks (DAAs), etc, which directly affect the QCR-A/B factors. The generation limits inequality condition of Eq. (23) is modified with Eqs.~\eqref{eq:35}--\eqref{eq:38}, while the other inequality conditions of the problem remain the same. The binary variable $\zeta$ determines if the generator $k$ must be disabled or adjusted according to HiTL preferences, e.g. if a generator is considered fully compromised and must be disconnected ($\zeta=1$) instead of just curtailed ($\zeta=0$).

As seen in the above formulation, $\rho$ defines the boundaries, which must be defined by experts operating the system, between a `reliable' and an `unreliable' generation node. If the CPES-QSM value (shown here as $CQ$) is greater than or equal to $\rho$ then the node is considered `unreliable' and its generation is either curtailed to a range of its minimum to maximum values or completely turned off. On the other hand, if $CQ$ is lower than $\rho$ then the node is considered `reliable' and there is no change in its maximum dispatch capacity. It is important to note that the curtailment of the generators can be defined as a continuous function that curtails generation according to a specific percentage value instead of $ P_k^{gl}$. By constraining the generation in `unreliable' nodes, the OPF solution provides a more `secure' solution that relies on the generation of more reliable nodes at the cost of more expensive generation that yields a higher traditional cost. The final C-ACOPF solution makes the system more secure in terms of the cyber-physical security of the CPES while sacrificing generation cost. 

\hl{In terms of complexity, it should be noted that the C-ACOPF problem is unchanged from the original ACOPF, as no new constraints are added. However, the feasible solution space for the optimization solver is reduced based on the newly added cyber-physical variables. The C-ACOPF is still a fully non-linear nonconvex formulation that is solved using a primal-dual interior-point method.}

%% file: 5-results.tex
\vspace{-2mm}
\section{Experimental Setup and Case Studies}

The code used to run the case studies presented in this section can be found in \cite{sourcegitlab}. The case study presented here is based on the IEEE RTS-24 test system, which is the first part of the IEEE RTS-96 test system presented in \cite{subcommittee1979ieee}. This system has 10 generators, 1 slack bus (bus \#12), 17 loads, and 32 transmission lines. The specific loads, lines, and generators capacities/parameters are taken from the case model `case24\_ieee\_rts.m' available in MATPOWER\cite{matpower}. For this case study, we assume that every physical node that exists in the electrical transmission network is mapped with a cyber node composed of just one IoT device at the cyber-layer of the CPES. This will greatly simplify the cybersecurity model of the system for explanation purposes since only the QCR-B factor will be used to characterize the vulnerabilities on the node; but, for more complex CPES, the QCR-A model can be used to model complex attack graph-based vulnerabilities that exist in multiple IoT devices. \\ 

\vspace{-8mm}
\subsection{Traditional ACOPF Results}
\begin{figure*}
\centering
  \noindent\makebox[\textwidth]{\includegraphics[width=15cm]{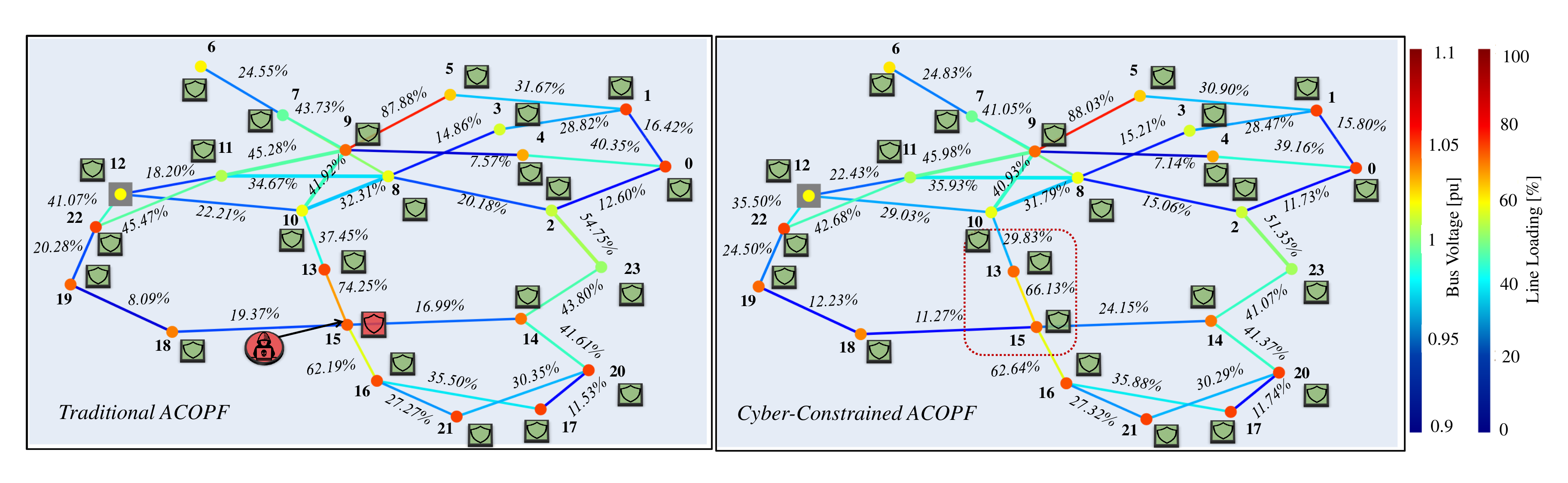}}
  \vspace{-2mm}
  \caption{Traditional ACOPF result (left) vs. Cyber-Constrained ACOPF (right) result for IEEE RTS-24 bus test system. The \textit{shields} represent the cyber-status of each node in the system. \textit{Green} means low QCR and \textit{red} means high QCR.}
  \vspace{-3mm}
  \label{fig:case1results}
\end{figure*}

In order to evaluate the advantages, in terms of the CPES security, of using the proposed C-ACOPF formulation, the solution of the T-ACOPF formulation is used as a baseline. The T-ACOPF solution is obtained by running the ACOPF PandaPower solver \cite{pandapower}, \hl{which uses the primal-dual interior-point method implemented using the Python Interior Point Solver (PIPS) in the PyPower package}. Table \ref{tab:case1tradacopf} shows the generator dispatch solution for the T-ACOPF formulation. The total cost for the solution is \$49,903.5432. Fig. \ref{fig:case1results} shows the bus voltages and line congestion of the solution.

\vspace{-2mm}

\begin{table}[t]
\setlength{\tabcolsep}{1.8pt}
\centering
\caption{Traditional ACOPF Results for IEEE RTS-24 Bus Test System.}
\label{tab:case1tradacopf}
\begin{tabular}{||c|c|c|c|c|c||}
\hline
\textbf{Gen \#} & \textbf{Bus \#} & \multicolumn{1}{l|}{\textbf{$P$ (MW)}} & \multicolumn{1}{l|}{\textbf{$Q$ (MVAR)}} & \multicolumn{1}{l|}{\textbf{$|V|_{pu}$}} & \multicolumn{1}{l||}{\textbf{$\angle V$}} \\ \hline \hline
0 & 0 & 192.00 & 13.42 & 1.050 & -7.38 \\ \hline
1 & 1 & 192.00 & 10.86 & 1.050 & -7.47 \\ \hline
2 & 6 & 131.60 & 66.68 & 1.022 & -17.84 \\ \hline
3 & 13 & 0.00 & 172.03 & 1.049 & 1.02 \\ \hline
4 & 14 & 215.0 & 110.00 & 1.042 & 10.03 \\ \hline
5 & 15 & 155.00 & 80.00 & 1.046 & 8.98 \\ \hline
6 & 17 & 400.00 & 69.02 & 1.050 & 14.83 \\ \hline
7 & 20 & 400.00 & -12.42 & 1.050 & 15.64 \\ \hline
8 & 21 & 300.00 & -39.00 & 1.050 & 21.27 \\ \hline
9 & 22 & 660.00 & 70.37 & 1.050 & 9.80 \\ \hline
\begin{tabular}[c]{@{}c@{}}10 \\ (slack)\end{tabular} & 12 & 258.54 & 53.05 & 1.020 & 0.00 \\ \hline
\end{tabular}
\vspace{-2mm}
\end{table}

\vspace{-3mm}

\subsection{Cyber-Constrained ACOPF using CPES-QSM}
In order to compute the C-ACOPF solution for the IEEE RTS-24 test system, we must first calculate the CPES-QSM metric value for each node in the system and `adjust' the corresponding generation and power flow of `unreliable' nodes identified based on the current status of the system. The first step in the process is to select the factors to be considered for this case study and assign the respective `expert' weights for each one of these factors. The factors selected for this case study are:  CRPI, QCR-B, VDI, SVSI, and VCPI. The `expert' weights, i.e., fuzzy measure values, assigned to each one of these factors are: $[0.26,0.55,0.61,0.65,0.66]$, respectively. A higher value represents higher individual importance of the factor in the final score computation. These weights were estimated after running multiple test cases and manually evaluating the operating conditions of the test system. Based on the number of factors used, there are $2^5=32$ total fuzzy measure values that go from  $\nu(\emptyset)=0.0$ to $\nu(\{x_1,x_2,x_3,x_4,x_5\}) =1.0$. Some of the intermediate values are $\nu(\{x_1,x_2\}) =0.669$, $\nu(\{x_1,x_3\}) =0.714$, and, $\nu(\{x_1,x_4\}) =0.743$. Some of them are not listed due to space limitations, but their calculation is trivial. The corresponding $\lambda$ computed is $\lambda=-0.983$.

Using the $\lambda$ and fuzzy measure values estimated, we can proceed to evaluate the current status of the operating CPES by calculating the CPES-QSM metric value for each node in the system. It is important to remark that at this point, the weights or fuzzy measure values are set and do not need to be recomputed every time the CPES-QSM metric value is estimated. Hence, for computing the CPES-QSM metric for each node, the current operating \textit{measurements} for each factor considered are obtained and processed to compute the corresponding CPES-QSM for that specific node. For this case study, an ACPF is run in order to estimate the current operating measurements for each node, which in turn are used to compute each factor considered. It is important to note that in a `real' system, these values are obtained either through IoTs and/or SE. Table \ref{tab:case1cpesrtsv} presents the calculated CPES-QSM values for each node in the system according to the corresponding factors computed. \hl{The time execution of the T-ACOPF is 1.1209 seconds, while the time execution of the entire CPES-QSM value computation and C-ACOPF is 1.3647 seconds. The specifications of the machine in which the tests were performed are: CPU - AMD 4900HS clocked at 3.00 GHz and 16 GB RAM. Based on the tests conducted, the entire CPES-QSM computation and C-ACOPF process seemed to be 0.82x slower, on average, than the T-ACOPF process.}

\begin{table}[t]
\setlength{\tabcolsep}{1.8pt}
\centering
\caption{CPES-QSM ($CQ$) Results for IEEE RTS-24 Bus Test System. Traditional (T-ACOPF) (referred as T) vs Cyber-Constrained ACOPF (C-ACOPF) (referred as C).}
\label{tab:case1cpesrtsv}
\begin{tabular}{||c||c|c||c|c||c|c||c|c||c|c||c|c||}
\hline
\textbf{\begin{tabular}[c]{@{}c@{}}Bus\\ \#\end{tabular}} & \multicolumn{2}{c|}{\textbf{CRPI}} & \multicolumn{2}{c|}{\textbf{QCR-B}} & \multicolumn{2}{c|}{\textbf{VDI}} & \multicolumn{2}{c|}{\textbf{SVSI}} & \multicolumn{2}{c|}{\textbf{VCPI}} & \multicolumn{2}{c||}{\textbf{$CQ$}} \\ \hline \hline
\textbf{Case} & \textbf{T} & \textbf{C} & \textbf{T} & \textbf{C} & \textbf{T} & \textbf{C} & \textbf{T} & \textbf{C} & \textbf{T} & \textbf{C} & \textbf{T} & \textbf{C} \\ \hline
0 & 0.11 & 0.11 & 0.0 & 0.0 & 0.05 & 0.05 & 0.0 & 0.0 & 0.0 & 0.0 & 0.05 & 0.05 \\ \hline
1 & 0.22 & 0.22 & 0.0 & 0.0 & 0.05 & 0.05 & 0.0 & 0.0 & 0.0 & 0.0 & 0.08 & 0.08 \\ \hline
2 & 0.12 & 0.12 & 0.03 & 0.03 & 0.01 & 0.01 & 0.02 & 0.0 & 0.01 & 0.01 & 0.05 & 0.04 \\ \hline
3 & 0.11 & 0.11 & 0.02 & 0.02 & 0.02 & 0.02 & 0.01 & 0.0 & 0.02 & 0.02 & 0.04 & 0.04 \\ \hline
4 & 0.11 & 0.11 & 0.01 & 0.01 & 0.03 & 0.03 & 0.01 & 0.01 & 0.01 & 0.01 & 0.05 & 0.05 \\ \hline
5 & 0.47 & 0.47 & 0.02 & 0.01 & 0.03 & 0.03 & 0.01 & 0.01 & 0.02 & 0.02 & 0.14 & 0.14 \\ \hline
6 & 0.09 & 0.09 & 0.01 & 0.01 & 0.05 & 0.05 & 0.0 & 0.0 & 0.03 & 0.03 & 0.04 & 0.04 \\ \hline
7 & 0.11 & 0.11 & 0.02 & 0.02 & 0.01 & 0.01 & 0.02 & 0.02 & 0.03 & 0.03 & 0.05 & 0.05 \\ \hline
8 & 0.12 & 0.12 & 0.03 & 0.03 & 0.02 & 0.02 & 0.01 & 0.01 & 0.01 & 0.01 & 0.04 & 0.04 \\ \hline
9 & 0.47 & 0.47 & 0.03 & 0.03 & 0.05 & 0.05 & 0.03 & 0.03 & 0.02 & 0.02 & 0.15 & 0.15 \\ \hline
10 & 0.19 & 0.19 & 0.03 & 0.03 & 0.02 & 0.02 & 0.01 & 0.01 & 0.02 & 0.02 & 0.07 & 0.07 \\ \hline
11 & 0.40 & 0.40 & 0.02 & 0.02 & 0.01 & 0.01 & 0.0 & 0.0 & 0.02 & 0.02 & 0.12 & 0.12 \\ \hline
12 & 0.53 & 0.53 & 0.01 & 0.01 & 0.02 & 0.02 & 0.0 & 0.0 & 0.0 & 0.0 & 0.15 & 0.15 \\ \hline
13 & 0.23 & 0.23 & 0.0 & 0.0 & 0.04 & 0.04 & 0.0 & 0.0 & 0.01 & 0.01 & 0.08 & 0.08 \\ \hline
14 & 1.0 & 1.0 & 0.02 & 0.0 & 0.04 & 0.04 & 0.0 & 0.0 & 0.0 & 0.0 & \cellcolor{red}0.27 & \cellcolor{red}0.27 \\ \hline
15 & 0.50 & 0.50 & \cellcolor{yellow}0.20 & \cellcolor{green}0.07 & 0.05 & 0.04 & 0.0 & 0.0 & 0.0 & 0.0 &\cellcolor{red}0.21 & \cellcolor{green}0.16 \\ \hline
16 & 0.41 & 0.41 & 0.02 & 0.02 & 0.05 & 0.05 & 0.03 & 0.03 & 0.0 & 0.0 & 0.13 & 0.13 \\ \hline
17 & 0.1 & 0.1 & 0.02 & 0.02 & 0.05 & 0.05 & 0.0 & 0.0 & 0.0 & 0.0 & 0.05 & 0.05 \\ \hline
18 & 0.50 & 0.50 & 0.02 & 0.02 & 0.04 & 0.04 & 0.03 & 0.02 & 0.0 & 0.0 & 0.15 & 0.15 \\ \hline
19 & 0.1 & 0.1 & 0.02 & 0.02 & 0.04 & 0.04 & 0.03 & 0.03 & 0.0 & 0.0 & 0.05 & 0.05 \\ \hline
20 & 0.31 & 0.31 & 0.04 & 0.04 & 0.05 & 0.05 & 0.0 & 0.0 & 0.0 & 0.0 & 0.10 & 0.10 \\ \hline
21 & 0.14 & 0.14 & 0.0 & 0.0 & 0.05 & 0.05 & 0.0 & 0.0 & 0.0 & 0.0 & 0.06 & 0.06 \\ \hline
22 & 0.53 & 0.53 & 0.02 & 0.02 & 0.05 & 0.05 & 0.0 & 0.0 & 0.01 & 0.01 & 0.16 & 0.16 \\ \hline
23 & 1.0 & 1.0 & 0.02 & 0.02 & 0.01 & 0.01 & 0.01 & 0.01 & 0.02 & 0.02 & \cellcolor{red}0.27 & \cellcolor{red}0.27 \\ \hline
\end{tabular}
\vspace{-2mm}
\end{table}

As seen in Table \ref{tab:case1cpesrtsv}, and using a $\rho=0.2$ value (where $\rho$ is the limit given by experts that provides the limit difference between a `reliable' node and an `unreliable' node), there are a few nodes that can be categorized as `unreliable`. This is due to the fact that the estimated CPES-QSM value (i.e, $CQ$) is higher than the limit given by the $\rho=0.2$, and in turn, some of the factors of these nodes tend to be higher than the rest, making them more susceptible to be targeted by attackers. `Case' \textbf{T} represents the case where the CPES-QSM value is estimated based on the T-ACOPF. `Case' \textbf{C} represents the case where the CPES-QSM value is estimated after the C-ACOPF is run and adjustments are made based on Eqs. (\ref{eq:35})-(\ref{eq:38}). It should be noted that the values in the table were cut to two significant figures due to space limitations. Results are normally presented using 5-6 significant figures, so further minor differences can be observed in some cases. For this case study, a DAA-type attack, e.g., Denial-of-Service (DoS), is considered as the threat to the system. The threat model for the cyberthreat is based on the threat model presented in \cite{cpessecurity}. This DAA threat has the capability of exploiting the vulnerabilities of the affected node(s) making them unresponsive (by delaying control and measurements) to operating commands. In our specific case, all nodes except bus \#15 (the affected node) are considered to have a very `secure' operation based on the QCR-B factor estimated. These are categorized using the following parameters \textit{\{AV: Local, PR: High, AC: High, UI: Required\}}, while bus \#15 is categorized as `unreliable' based on \textit{\{AV: Network, PR: None, AC: Low, UI: None\}} vulnerabilities exploited by the threat. The results obtained showcase how the cyber factors are taken into consideration when deciding the optimal AC power flow of a CPES that is currently under attack from a cyber perspective. Table \ref{tab:case1cyberacopf} shows the generator dispatch solution for the C-ACOPF formulation. The total cost for the solution is \$53,621.1357.

Fig. \ref{fig:case1results} shows the bus voltages and line congestion of the C-ACOPF solution. As seen in the figure, the OPF solution gets adjusted to take into account the `unreliable' nature of bus \#15 (currently under attack), which in turn derives from the high cyber and physical factors. The power flows from bus \#15 to bus \#13 and from bus \#15 to \#18  get reduced by $\approx$ 8\%, while the power flows from bus \#14 to bus \#15 and from bus \#16 to bus \#15 increase by $\approx$ 7\% and 0.45\%, respectively. The corresponding CPES-QSM values of the cyber-constrained solution are shown in the last column of Table \ref{tab:case1cpesrtsv}. As seen, the $CQ$ value for node \#15 (affected) decreased from 0.21 to 0.16, putting it below the $\rho$ threshold. These results clearly demonstrate how the C-ACOPF increases the generation from other more `reliable' sources and takes into account the current status of not only the physical system but the cyber-system components; thus, producing results that make the system more secure and resilient from the cybersecurity point of view. It is important to note how the CPES-QSM values for nodes \#14 and \#23 are still over the $\rho$ threshold; however, these cannot be directly adjusted based on the ACOPF solution since their criticality is derived from the CRPI factor, which in turn, is related to the physical topology of the system. This factor can be reduced by performing other control operations that directly change the topology of the system, but these control operations are not in the scope of this paper. Future work will focus on exploring other control solutions that can be adjusted based on the CPES-QSM metric.

\begin{table}[t]
\setlength{\tabcolsep}{1.8pt}
\centering
\caption{Cyber-Constrained ACOPF Results for IEEE RTS-24 Bus Test System}
\label{tab:case1cyberacopf}
\begin{tabular}{||c|c|c|c|c|c||}
\hline
\textbf{Gen \#} & \textbf{Bus \#} & \multicolumn{1}{l|}{\textbf{$P$ (MW)}} & \multicolumn{1}{l|}{\textbf{$Q$ (MVAR)}} & \multicolumn{1}{l|}{\textbf{$|V|_{pu}$}} & \multicolumn{1}{l||}{\textbf{$\angle V$}} \\ \hline \hline
0 & 0 & 192.00 & 12.55 & 1.050 & -8.33 \\ \hline
1 & 1 & 192.00 & 10.54 & 1.050 & -8.40 \\ \hline
2 & 6 & 141.64 & 64.54 & 1.024 & -17.76 \\ \hline
3 & 13 & 0.00 & 146.10 & 1.044 & -0.79 \\ \hline
4 & 14 & 215.0 & 110.00 & 1.042 & 7.47 \\ \hline
5 & 15 & \cellcolor{green}54.30 & 80.00 & 1.044 & 6.32 \\ \hline
6 & 17 & 400.00 & 73.54 & 1.050 & 12.22 \\ \hline
7 & 20 & 400.00 & -10.61 & 1.050 & 13.05 \\ \hline
8 & 21 & 300.00 & -38.39 & 1.050 & 18.67 \\ \hline
9 & 22 & 660.00 & 68.44 & 1.050 & 8.40 \\ \hline
\begin{tabular}[c]{@{}c@{}}10 \\ (slack)\end{tabular} & 12 & 344.70 & 43.15 & 1.021 & 0.00 \\ \hline 
\end{tabular}
\vspace{-2mm}
\end{table}

\vspace{-2mm}
\subsection{Effects of Cyberattacks in Traditional ACOPF and Cyber-Constrained ACOPF Formulations}

In this subsection, the effects that a cyberattack may have in a system, as the one evaluated in the previous subsection, are explored. The effects are evaluated by comparing the frequency and voltage effects of the system when a cyberattack compromises a vulnerable node. For this case study, we use the dispatch results from the T-ACOPF and the C-ACOPF in order to examine the different responses these systems would have when a DAA threat (capable of exploiting the vulnerabilities of the affected node(s) by making them unresponsive via the delay of control and measurements) is deployed. More specifically, in this case, the cyber threat compromises the vulnerable generator at the `unreliable' bus identified by the C-ACOPF formulation, i.e., generator at bus \#15, by making it inoperable for 5 seconds. The threat model of such an attack is similar to existing attacks found in the literature \cite{xenofontos2021, cpessecurity}. The time-domain simulation of the IEEE RTS-24 test system used for this analysis is performed using the Power System Analysis Toolbox (PSAT)\cite{milano2008open}.

\begin{figure}[t]
\centering
\includegraphics[width = 0.32\textwidth]{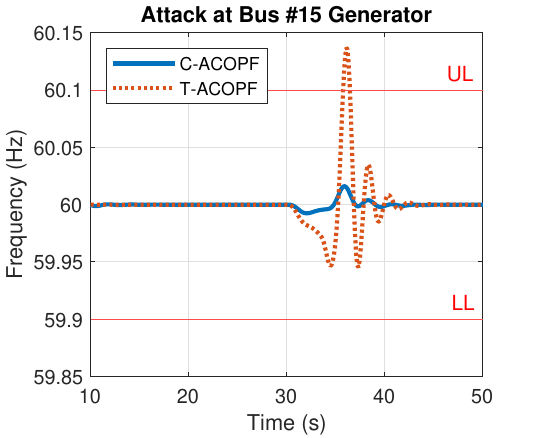}
\vspace{-2mm}
\caption{\label{fig:freqcomp} Frequency response comparison for both T-ACOPF and C-ACOPF solutions when a $5$ seconds cyberattack is used to compromise the generator at bus \#15. (UL means \textit{upper limit} while LL means \textit{lower limit}.)}
\vspace{-2mm}
\end{figure}

\begin{figure}[t]
\centering
\includegraphics[width = 0.32\textwidth]{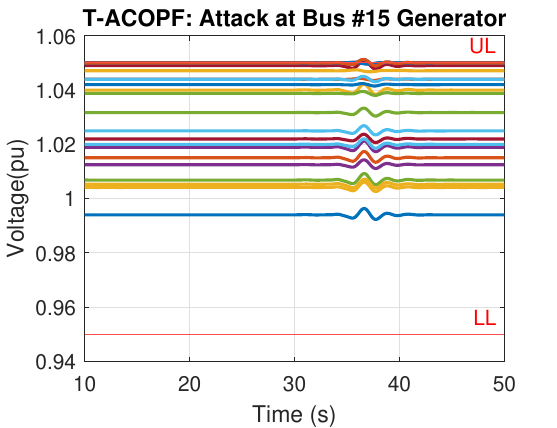}
\vspace{-2mm}
\caption{\label{fig:volt_tacopf} Voltage values (in pu) for T-ACOPF solution when a $5$ seconds cyberattack is used to compromise the generator at bus \#15. Voltage of all buses in the system are represented by different color lines. (UL means \textit{upper limit} while LL means \textit{lower limit}.)}
\vspace{-2mm}
\end{figure}

\begin{figure}[t]
\centering
\includegraphics[width = 0.32\textwidth]{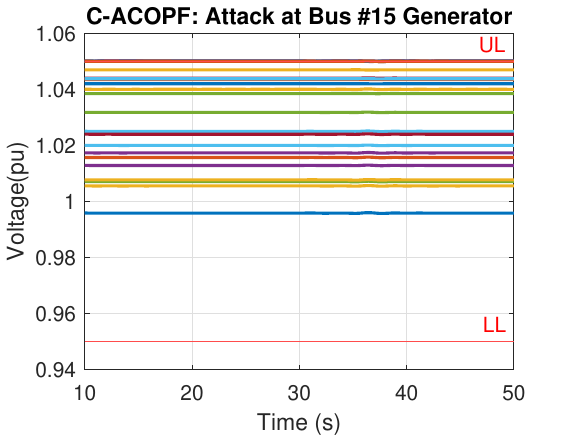}
\vspace{-2mm}
\caption{\label{fig:volt_cacopf} Voltage values (in pu) for C-ACOPF solution when a $5$ seconds cyberattack is used to compromise the generator at bus \#15. Voltage of all buses in the system are represented by different color lines. (UL means \textit{upper limit} while LL means \textit{lower limit}.)}
\vspace{-2mm}
\end{figure}

Fig. \ref{fig:freqcomp} showcases the differences in the frequency response for both the T-ACOPF and C-ACOPF solutions when a 5 seconds DAA is used to compromise the IoT(s) connected to the generator at bus \#15. The cyberattack was carried out at $t$=30s to $t$=35s. As seen in the figure, the frequency variation of the T-ACOPF solution system is significantly higher than the frequency variation caused by the cyberattack in the C-ACOPF solution system.  The frequency even crosses for an instance the upper limit (at 60.1 Hz) of the system's frequency. Figs. \ref{fig:volt_tacopf} and \ref{fig:volt_cacopf} depict the voltages for all the nodes in the system for the time period evaluated. As observed in these figures, the voltage variations are also much more notable in the T-ACOPF system when compared to the C-ACOPF system. In fact, in the C-ACOPF solution system, the cyberattack is barely noticeable, in terms of voltage values.

These results clearly demonstrate that the use of the proposed C-ACOPF formulation, enhanced by the CPES-QSM, has major advantages when compared to the T-ACOPF formulation due to the fact that this new formulation and metric consider not only the physical state of the system but also the current cyber state in order to improve control and operation decisions that could make the system more secure and stable. In this case study, the C-ACOPF correctly cataloged bus \#15 as an `unreliable' node due to its high QCR-B value and limited its generation output to a minimum. So, when the node got compromised (attacked by an adversary), the system did not suffer much and was able to endure any possible subsequent damage. Since the T-ACOPF solution was not `cyber aware' as the C-ACOPF solution, the system presented a higher variability when the attack happened and a longer attack (e.g., 15+ seconds) may have caused more substantial damage. The loss of 54.30 MW generation is not the same as the loss of 155 MW generation and this is reflected in the results.  

%% file: 6-conclusion.tex
\section{Conclusions}
In this paper, a quantitative cyber-physical security metric for CPES and a cyber-constrained ACOPF formulation are proposed to cope with the security challenges of modern CPES. The CPES-QSM cyber-metric is designed to incorporate diverse types of critical factors affecting the cybersecurity and operation of CPES while providing a quantitative value to the cyber and physical status of the operating CPES. This cyber-metric is integrated, as a proxy, to transform the traditional ACOPF formulation into a cyber-constrained ACOPF formulation designed to produce a more secure operating point that considers vulnerabilities existing in IoT and OT devices deployed in the system's physical and cyber networks. Experimental case studies, based on a DAA-type of attack and using the IEEE RTS-24 bus system, are presented to show the effectiveness of the proposed CPES cyber-metric. The results of the case study are discussed and evaluated against the traditional ACOPF formulation, in terms of cost, line loading percentage, and CPES security. 

\hl{Based on tests conducted, some of the factors considered when computing the CPES-QSM may have a significant impact in the scalability of the proposed method. Therefore, future work will focus on exploring the scalability of the proposed approach by evaluating its performance in large-scale integrated transmission-distribution systems using tools such as \textit{PowerModelsITD.jl}\footnote{https://github.com/lanl-ansi/PowerModelsITD.jl}. Furthermore, the stability of the CI will be examined for the case when a large number of factors are considered simultaneously.}